\documentclass[12pt]{article}
\pdfoutput=1
\usepackage{amsfonts,hyperref}
\usepackage{color}
\usepackage{appendix}
\usepackage{amsmath}
\usepackage{graphicx}
\usepackage[latin1]{inputenc}
\usepackage[french,english]{babel}
\usepackage{geometry}
\usepackage{etoolbox}
\usepackage{fancyhdr}
\patchcmd{\thebibliography}{\section*}{\section}{}{}

\addto\captionsfrench{}
\addto\captionsenglish{}

\newcommand\encadremath[1]{\vbox{\hrule\hbox{\vrule\kern8pt
\vbox{\kern8pt \hbox{$\displaystyle #1$}\kern8pt}
\kern8pt\vrule}\hrule}}
\def\enca#1{\vbox{\hrule\hbox{
\vrule\kern8pt\vbox{\kern8pt \hbox{$\displaystyle #1$}
\kern8pt} \kern8pt\vrule}\hrule}}

\newcommand\figureframex[3]{
\begin{figure}[bth]
\hrule\hbox{\vrule\kern8pt
\vbox{\kern8pt \vbox{
\begin{center}
{\mbox{\epsfxsize=#1.truecm\epsfbox{#2}}}
\end{center}
\caption{#3}
}\kern8pt}
\kern8pt\vrule}\hrule
\end{figure}
}

\makeatletter
\@addtoreset{equation}{section}
\makeatother
\newtheorem{theorem}{Theorem}[section]
\newtheorem{conjecture}{Conjecture}[section]
\newtheorem{remark}{Remark}[section]
\newtheorem{proposition}{Proposition}[section]
\newtheorem{lemma}{Lemma}[section]
\newtheorem{corollary}{Corollary}[section]
\newtheorem{definition}{Definition}[section]

\def\br{\begin{remark}\rm\small}
\def\er{\end{remark}}
\def\bt{\begin{theorem}}
\def\et{\end{theorem}}
\def\bd{\begin{definition}}
\def\ed{\end{definition}}
\def\bp{\begin{proposition}}
\def\ep{\end{proposition}}
\def\bl{\begin{lemma}}
\def\el{\end{lemma}}
\def\bc{\begin{corollary}}
\def\ec{\end{corollary}}
\def\beaq{\begin{eqnarray}}
\def\eeaq{\end{eqnarray}}

\newcommand{\beq}{\begin{equation}}
\newcommand{\eeq}{\end{equation}}
\newcommand{\bea}{\begin{eqnarray}}
\newcommand{\eea}{\end{eqnarray}}
\newcommand{\beqq}{\begin{equation*}}
\newcommand{\eeqq}{\end{equation*}}
\newcommand{\beaa}{\begin{eqnarray*}}
\newcommand{\eeaa}{\end{eqnarray*}}

\newcommand{\Pint}{{\int\kern -1.em -\kern-.25em}}

\begin{document}

\sloppy


\pagestyle{empty}
\vspace{10pt}
\begin{center}
{\large \bf {Elements of proof for conjectures of Witte and Forrester about the combinatorial structure of Gaussian Beta Ensembles}}
\end{center}
\vspace{5pt}
\begin{center}
\textbf{O. Marchal$^\dagger$}
\end{center}
\vspace{20pt}

$^\dagger$ \textit{Université de Lyon, CNRS UMR 5208, Université Jean Monnet, Institut Camille Jordan, France}
\footnote{olivier.marchal@univ-st-etienne.fr}

\vspace{30pt}

{\bf Abstract}:
The purpose of the article is to provide partial proofs for two conjectures given by Witte and Forrester in ``Moments of the Gaussian $\beta$ Ensembles and the large $N$ expansion of the densities" (\cite{WF}) with the use of the topological recursion adapted for general $\beta$ Gaussian case. In particular, the paper uses a version at coinciding points that provides a simple proof for some of the coefficients involved in the conjecture. Additionally, we propose a generalized version of the conjectures for all correlation functions evaluated at coinciding points.


\vspace{30pt}
\pagestyle{plain}
\setcounter{page}{1}


\section{Notation and loop equations}

\subsection{Introduction to the problem}

In this section we remind the general formalism (for simplicity, we try to keep the same notation as the one developed in \cite{WF}) used to write the loop equations for the Gaussian Beta ensembles. The Gaussian Beta Ensembles are defined by the following partition function:
\beq \label{BetaMM} Z_N=\int d\lambda \prod_{i<j} |\lambda_i-\lambda_j|^{2\kappa} e^{-\frac{N\kappa}{T}
\underset{i=1}{\overset{N}{\sum}} \frac{\lambda_i^2}{2}}\eeq
We will define $\kappa=\frac{\beta}{2}$ to match the convention of \cite{WF}. Note that our coupling constant is written $T$ instead of $g$ as used in \cite{WF}. The potential is quadratic and is given by $V(x)=\frac{x^2}{2}$. The hermitian case is recovered as usual for $\kappa=1$ (or $\beta=2$ with our convention). We also define the free energy (note the conventional minus sign in the following definition):
\beq F=-\ln Z_N\eeq 
The general purpose in the study of matrix models is to determine the large $N$ asymptotic of integrals of the form \eqref{BetaMM}. In the case of hermitian matrix models (and to some extent for $\beta=1$ or $\beta=4$), there are several methods to obtain it but most of them fail to extend to the general $\beta$ case. For example, the orthogonal polynomials strategy \cite{Mehta} has not been generalized to the general $\beta$ case so far. Other method like tridiagonalization of the matrix by the householder algorithm and connection with stochastic differential equations \cite{RRV, ABG, AD} are possible but become rapidly inefficient to get subleading orders of the large $N$ expansion. However universality results for local statistics (universal in the sense that they do not depend on the potential $V(x)$) have been proved to hold for general beta using and therefore it appears important to understand better the Gaussian case for which results are usually easier to derive. The strategy used in \cite{WF} is to solve the so-called ``loop equations" by using an adaptation of the topological recursion for general beta. Historically, the topological recursion has been developed to solve the hermitian matrix models loop equations and was later adapted to any spectral curve. This situation is nicely understood (See \cite{EO} for the general theory) and many geometric or combinatorial identities have been derived (or re-derived) with this formalism. The situation is different regarding $\beta$-ensembles. Indeed, as we will see later, loop equations for $\beta$-ensembles can be derived in the same way as the hermitian case but an important simplification arises only when $\beta=2$. Solving the loop equations for arbitrary $\beta$ and a general potential $V(x)$ still remains an open question although some attempts were tried \cite{BMS, M, BM, CBM, CBM2}. In our case, since the model is Gaussian the situation is much easier and the loop equations can be solved recursively. However as we will see the recursive solution is mostly formal since the computations rapidly become tedious and hide the combinatorial aspects of the correlations functions. In \cite{WF} some conjectures were proposed and we plan to prove pieces of them in this paper. Eventually we believe that understanding properly the Gaussian case is essential since recent universality results (in the bulk or at the edge of the spectrum) \cite{Bour, Bour2, BG, BG2} prove that the Gaussian case can be used to describe local statistics arising for any potential $V(x)$. We will now introduce the correlation functions, the large $N$ expansion and remind the conjecture proposed by Witte and Forrester.
\medskip

\subsection{Correlation functions}

For an arbitrary potential $V(x)$, it is standard to define the following functions:
\bea
W_n(x_1,\dots,x_n)&=&\left<\sum_{i_1,\dots,i_n=1}^N \frac{1}{x_1-\lambda_{i_1}}\dots \frac{1}{x_n-\lambda_{i_n}}\right>_c\cr
Q_{n+1}(x;x_1,\dots,x_n)&=&\left<\sum_{i,i_1,\dots,i_n=1}^N \frac{V'(x)-V'(\lambda_{i})}{x-\lambda_i}\frac{1}{x_1-\lambda_{i_1}}\dots\frac{1}{x_n-\lambda_{i_n}}\right>_c\cr
\eea
with the convention that $Q_1(x)=\left<\underset{i=1}{\overset{N}{\sum}}\frac{V'(x)-V'(\lambda_{i})}{x-\lambda_i}\right>$. The index $c$ stands for the ``connected'' or ``cumulant'' part. As usual, the bracket notation corresponds to taking the average relatively to the measure defined by \eqref{BetaMM}. Note that in the definition of $P_{n+1}$, the cumulant part only applies to the last $n^\text{th}$ variables but not to the first one. The functions $W_n(x_1,\dots,x_n)$ are known as correlation functions. In particular $W_1(x)$ is usually called the resolvent or the one-point function. Moreover, in many articles the notation is extended to $n=0$ by defining $W_0=F=-\ln Z_N$. When $V(x)$ is polynomial, the functions $P_{n+1}(x;x_1,\dots,x_n)$ are also polynomial functions in their first variable, a key observation to solve the topological recursion in the hermitian case.

\medskip

In the Gaussian case, the situation simplifies greatly because we have:
\bea Q_1(x)&=&N\cr
Q_{n+1}(x,x_1,\dots,x_n)&=&0 \,\,,\,\, \forall n\geq 1\cr
\eea
Indeed, $V'(x)=x$ and therefore $Q_1(x)=\left<\underset{i=1}{\overset{N}{\sum}}\, 1\right>=N$. For $n\geq 1$, $Q_{n+1}$ vanishes because the cumulant part makes it zero by symmetry.

\medskip

\subsection{Loop equations}
It is known for general potentials $V(x)$ (See \cite{Eyn}) that the previous functions satisfy the so-called loop equations (we denote $I=\{x_1,\dots,x_{n-1}\}$):

\beq \label{FirstLoop} W_1(x)^2-\frac{N}{T}V'(x)W_1(x)+\left(1-\frac{1}{\kappa}\right)W_1'(x)+W_2(x,x)+\frac{N}{T}Q_1(x)=0\eeq

and $\forall n>1$:
\bea \label{GeneralLoop} &&\left[\frac{NV'(x)}{T}-2W_1(x)+\left(\frac{1}{\kappa}-1\right)\partial_x\right]W_n(x,I)=W_{n+1}(x,x,I)\cr
&&+\sum_{J\subset I, J\notin \{\emptyset,I\} } W_{|J|+1}(x,J)W_{n-|J|}(x,I\setminus J)+\frac{N}{T}Q_n(x;I)+\frac{1}{\kappa}\sum_{x_i\in I}\frac{\partial}{\partial x_i}\frac{W_{n-1}(x,I\setminus\{x_i\})-W_{n-1}(I)}{x-x_i}\cr
\eea
These equations can be easily derived with infinitesimal transformations or with a suitable integration by parts. A direct observation shows that the case $\kappa=1$ (i.e. $\beta=2$) is special since the coefficient in front of $\partial_x W_n(x,I)$ vanishes.

\subsection{Large $N$ expansion}

In many cases, it can be proved that the correlation functions have a series expansion in $\frac{1}{N}$ at large $N$. In particular this situation is expected when the potential $V(x)$ has only one minimum around which the eigenvalues are expected to accumulate. In other cases, the situation may not be as nice and the $\frac{1}{N}$ expansion is also known not to reproduce the complete asymptotic expansion of the correlation functions (See \cite{BDE, BG2}). We will not say more about these aspects here and refer the reader to the standard literature about this issue. We just mention here that this issue is mostly irrelevant for the Gaussian case since it has been proved recently \cite{BG} that in the Gaussian case there exists a large $N$ expansion of the form:
\bea \label{NExpansion}
W_n(x_1,\dots,x_n)&=&\sum_{l=0}^\infty \left(\frac{N}{T}\right)^{2-n-l}\sqrt{\kappa}^{2-2n-l} W_n^{(l)}(x_1,\dots,x_n)\cr
Q_n(x;x_1,\dots,x_{n-1})&=&\sum_{l=0}^\infty \left(\frac{N}{T}\right)^{2-n-l}\sqrt{\kappa}^{2-2n-l} Q_n^{(l)}(x_1,\dots,x_n)\cr
F&=&\sum_{l=0}^\infty \left(\frac{N\sqrt{\kappa}}{T}\right)^{2-l}F^{(l)}\cr
\eea
In particular we have both even and odd powers of $N$ in the expansion whereas in the hermitian case only even powers appear. The structure of the loop equations implies that each $W_n^{(l)}$ is a polynomial in $\left(\sqrt{\kappa}-\frac{1}{\sqrt{\kappa}}\right)$ of degree $l$ with the upper half of its coefficients only determined by $V(x)$.

\section{Solving recursively the loop equations in the Gaussian case}

\subsection{Initialization: spectral curve}
We now focus on the Gaussian case and we project the loop equation \eqref{FirstLoop} to its leading order in $\frac{1}{N}$. We get:
\beq W_1^{(0)}(x)=\frac{1}{2}\left(x-\sqrt{x^2-4T}\right)\eeq
which gives the standard semi-circular law. We introduce the following notation:
\beq \label{y} y(x)=V'(x)-2W_1^{(0)}(x)=\sqrt{x^2-4T} \,\,,\,\, y^2(x)=x^2-4T\eeq
Note that this equation is independent of $\kappa$ and therefore corresponds to the usual spectral curve of the Gaussian Hermitian matrix model. 

\subsection{Rewriting of the loop equations}
In order to have more compact notations we define:
\beq \label{hbar} \hbar\overset{\text{def}}{=}\sqrt{\kappa}-\frac{1}{\sqrt{\kappa}}\eeq
Let us now project the loop equations \eqref{FirstLoop} to the various powers of the series expansion in $\frac{1}{N}$. Remembering that the functions $Q_1^{(g)}(x)$ vanish for $g>0$ we get: 

\beq \label{W11Loop} y(x)W_1^{(1)}(x)= \hbar\partial_x W_1^{(0)}(x)\eeq
and $\forall g\geq 2:$
\beq \label{ExpandedFirstLoop} 
y(x)W_1^{(g)}(x)=\sum_{p=1}^{g-1}W_1^{(p)}(x)W_1^{(g-p)}(x)+\hbar\partial_x W_1^{(g-1)}(x)+W_2^{(g-2)}(x,x)
\eeq
Equation \eqref{W11Loop} gives $W_1^{(1)}(x)$ while \eqref{ExpandedFirstLoop} taken at $g=0$ gives $W_2^{(0)}(x,x)$ but not the complete function $W_2^{(0)}(x_1,x_2)$. Note that equation \eqref{ExpandedFirstLoop} is only useful to determine $W_1^{(g+2)}(x)$ from the knowledge of $W_2^{(g)}(x,x)$ but the opposite way would not determine $W_2^{(g)}(x_1,x_2)$ but only its diagonal part $W_2^{(k)}(x,x)$.

The expansion of \eqref{GeneralLoop} gives the following set of equations ($n\geq 1$ and $g\geq 0)$:

\bea \label{GeneralLoopExpanded} &&y(x)W_n^{(g)}(x,I)=\hbar\partial_x W_n^{(g-1)}(x,I)+W_{n+1}^{(g-2)}(x,x,I)\cr
&&+\sum_{J\subseteq I}'\sum_{p=0}^g W_{|J|+1}^{(p)}(x,J)W_{n-|J|}^{(g-p)}(x,I\setminus J)+\sum_{x_i \in I}\frac{\partial}{\partial x_i}\left(\frac{W_{n-1}^{(g)}(x,I\setminus x_i)-W_{n-1}^{(g)}(I)}{x-x_i}\right)\cr
\eea 

with the notation $'$ on the double sum indicating that the terms $(J=\emptyset,p=0)$ and $(J=I,p=g)$ should be discarded. Moreover we use here the convention that $W_n^{(-1)}$ and $W_n^{(-2)}$ are identically zero. We clearly see that the equations only involve $\hbar=\sqrt{\kappa}-\frac{1}{\sqrt{\kappa}}$ but not directly $\kappa$ itself. In particular it means that we have the symmetry $\kappa \rightarrow \frac{1}{\kappa}$, a well-known fact in the matrix models literature. Finally, we observe that \eqref{ExpandedFirstLoop} is a special case of \eqref{GeneralLoopExpanded} with $n=1$ so that we can conveniently regroup them under the same notation.

\subsection{\label{Derivy} Observation for the derivatives of $y(x)$}
From the definition of the function $y(x)$ \eqref{y}, we have:
\beq y'(x)=\frac{x}{y(x)} \,\,,\,\, y''(x)=-\frac{4T}{y(x)^3}\,\,,\,\, y^{(3)}(x)=\frac{12\,T x}{y(x)^5}\eeq 
and more generally:
\beq \forall\, n\geq 2\,:\,  y^{(n)}(x)=\frac{R_{n}(x)}{y(x)^{(2n-1)}} \text{ with } R_n(x) \text{ a polynomial of degree n-2 }  \eeq
The polynomials $R_n(x)$ satisfy the following recursion:
\beq \forall\,  n\geq 1: \,R_{n+1}(x)=(x^2-4T)R_n'(x)-(2n-1)xR_n(x) \text{ with } R_1(x)=x\eeq
Note that we have a special case when passing from $R_1(x)$ to $R_2(x)$ because the leading term cancels (that is why the degree of $R_2(x)$ is $0$ and not $2$). It is then straightforward to prove by induction that:
\begin{enumerate} 
\item $\forall \, n\geq 2\, :\, R_n(x)$ is a polynomial of degree $n-2$
\item $R_n(x)$ is even when $n$ is even and $R_n(x)$ is odd when $n$ is odd
\item The leading coefficient of $R_n(x)$ is given by $2T (-1)^{n+1} n!$
\item If we denote $R_n(x)=\underset{k=0}{\overset{n-2}{\sum}} a_k^{(n)}x^k$ the coefficients of the polynomials, then the non-zero coefficients are determined by:
\beq \forall \, 0\leq k\leq n-1\,:\, a_k^{(n+1)}=(k-2n)a_{k-1}^{(n)}-4T(k+1)a_{k+1}^{(n)}\eeq
with the convention that $a_{-1}^{(n)}=0$ and $a_{n-1}^{(n)}=0$ 
\end{enumerate} 
These properties are useful to compute the first correlation functions more efficiently.

\subsection{Computation of the first correlation functions}
In the Gaussian case the loop equations \eqref{ExpandedFirstLoop} and \eqref{GeneralLoopExpanded} can be solved recursively because they do not imply any unknown functions (that would not be the case for a general potential where some unknown $Q_n^{(g)}$ functions would appear). In order to illustrate the conjectures proposed in \cite{WF} we present here the first correlation functions. \textbf{In the rest of the paper we will denote $y$ for $y(x)$ and $y_i$ for $y(x_i)$ to shorten formulas}:
\beq  W_1^{(1)}(x)=\frac{\hbar}{2}\left(\frac{1}{y}-\frac{x}{y^2}\right) \eeq 
\beq W_2^{(0)}(x_1,x_2)=-\frac{1}{2(x_1-x_2)^2}+\frac{x_1x_2-4T}{2(x_1-x_2)^2y_1y_2}= -\frac{y_1y_2-x_1x_2+4T}{2(x_1-x_2)^2y_1y_2} \eeq
We observe that if we introduce $f(x_1,x_2)=x_1x_2-4T$ we get $y=f(x,x)$ so that $W_2^{(0)}(x_1,x_2)$ is regular at $x_1=x_2$ and we obtain:
\beq W_2^{(0)}(x,x)=\frac{T}{y^4} \eeq
From the loop equations, it is also easy to observe that the correlation functions $W_n^{(0)}(x_1,\dots,x_n)$ do not depend on $\kappa$ and are therefore identical to the standard hermitian ones. The next orders are:

\bea W_1^{(2)}(x) &=&\hbar^2\left[-\frac{x}{y^4}+\frac{x^2+T}{y^5}\right]+\frac{T}{y^5} \cr
W_1^{(3)}(x)&=&5\hbar^3\left(\frac{x^2+T}{y^7}- \frac{x(x^2+2T)}{y^8}\right)+\hbar\left(\frac{x^2+6T}{2y^7}-\frac{x(x^2+30T)}{2y^8}\right)\cr 
W_3^{(0)}(x_1,x_2,x_3)&=&\frac{2T(x_1x_2+x_1x_3+x_2x_3+4T)}{y_1^3y_2^3y_3^3} \cr
W_3^{(0)}(x,x,x)&=&\frac{2T(3x^2+4T)}{y^9} \eea
and
\bea &&W_4^{(0)}(x_1,x_2,x_3,x_4)=\frac{1}{y_1^5y_2^5y_3^5y_4^5}\Big[ -12288\,T^6\cr
&&T^5\Big(1536\,(x_1^2+\text{perm})-4096\,(x_1x_2+\text{perm})\Big)\cr
&&+T^4\Big(-1536\,x_1x_2x_3x_4+640\,(x_1x_2x_3^2+\text{perm})+640\,(x_1^3x_2+\text{perm})\Big)\cr
&&+T^3\Big(288\,(x_1x_2x_3x_4^3+\text{perm})-64\,(x_1x_2^2x_3^3+\text{perm})-64\,(x_1^3x_2^3+\text{perm})-64\,(x_1x_2x_3^2x_4^2+\text{perm})\cr
&&-96\,(x_1^2x_2^2x_3^2+\text{perm})\Big)\cr
&&+T^2\Big(48\,x_1^2x_2^2x_3^2x_4^2-48\,(x_1^3x_2^3x_3x_4+\text{perm})-8\,(x_1^3x_2^3x_3^2+\text{perm})-8\,(x_1^3x_2^2x_3^2x_4+\text{perm})+\Big)\cr
&&+T\Big(8\,(x_1^3x_2^3x_3^2x_4^2+\text{perm})+6\,(x_1^3x_2^3x_3^3x_4 +\text{perm})\Big) \Big]
\eea
The word ``perm" indicates that we include all other terms needed to obtain a symmetric polynomial in $(x_1,x_2,x_3,x_4)$. The evaluation at coinciding points is:
\bea W_4^{(0)}(x,x,x,x)&=&\frac{24T(3x^4+18\,Tx^2+8T^2)}{y^{14}}\cr
W_1^{(4)}(x)&=&\hbar^4\left(-\frac{37x^3+92Tx}{y^{10}}+\frac{37x^4+123Tx^2+21T^2}{y^{11}}\right)\cr
&&+\hbar^2\left(-\frac{23x^3+180Tx}{2y^{10}}+\frac{23x^4+454Tx^2+176T^2}{2y^{11}}\right)+\frac{21T(x^2+T)}{y^{11}}\cr
W_2^{(1)}(x,x)&=&\hbar\left(-\frac{x(x^2+18T)}{2y^7}+\frac{x^2+4T}{2y^6}\right)\cr
W_2^{(2)}(x,x)&=&\frac{T(20T+21x^2)}{y^{10}}+\hbar^2\left( \frac{98Tx^2+38T^2+8x^4}{y^{10}}- \frac{8x^3+45Tx}{y^9}\right)\cr\eea

We recover here the results presented in \cite{WF} in which the authors proposed the following conjecture:

\begin{conjecture} \label{Conj} (From \cite{WF}, page $9$) For $l\leq 2$ even we have:
\beq W_1^{(l)}(x)=\hbar^l\left[\frac{P_1^{l}(x)}{y^{3l-2}}+\frac{P_2^{l}(x)}{y^{3l-1}}\right]+\hbar^{l-2}\left[\frac{P_3^{l}(x)}{y^{3l-2}}+\frac{P_4^{l}(x)}{y^{3l-1}}\right]+\dots+\hbar^2 \left[\frac{P_{l-1}^{l}(x)}{y^{3l-2}}+\frac{P_l^{l}(x)}{y^{3l-1}}\right]+\frac{P_{l+1}^l(x)}{y^{3l-1}}\eeq
where $\text{deg}_x P_j^l=l-1$ for $j=1,3,\dots,l-1$ and $\text{deg}_x P_j^l=l$ for $j=2,4,\dots,l$ and $\text{deg}_x P_{l+1}^l=l-2$.
For $l\geq 1$ odd we have:
\beq W_1^l(x)=\hbar^l\left[\frac{P_1^{l}(x)}{y^{3l-2}}+\frac{P_2^{l}(x)}{y^{3l-1}}\right]+\hbar^{l-2}\left[\frac{P_3^{l}(x)}{y^{3l-2}}+\frac{P_4^{l}(x)}{y^{3l-1}}\right]+\dots +\hbar \left[\frac{P_l^{l}(x)}{y^{3l-2}}+\frac{P_{l+1}^{l}(x)}{y^{3l-1}}\right]\eeq
where the polynomials have $\text{deg}_x P_j^l=l-1$ for $j=1,3,\dots,l$ and $\text{deg}_x P_j^l=l$ for $j=2,4,\dots,l+1$. Moreover, the polynomials involved in both formulas are either even or odd functions of $x$ according to their degree. Furthermore, the leading term in the $x\to \infty$ expansion of $W_1^l(x)$ is of order $x^{-2l-1}$ for all $l\geq 0$.
\end{conjecture}

In their article, the authors presented computations up to $l=6$ supporting their conjecture. In this article we prove that the conjecture holds for certain coefficients and we propose a generalization to all correlation functions $W_n^{(l)}(x_1,\dots,x_n)$ evaluated at coinciding points $x_1=\dots=x_n$.

\section{Main theorem and generalization of the conjecture}

First from the loop equations \eqref{GeneralLoopExpanded} a straightforward induction shows that the functions $W_n^{(g)}(x_1,\dots,x_n)$ are polynomials in $\hbar$ of degree $g$ and are either even or odd functions of $\hbar$ relatively to their degree. Therefore we introduce the following definition:

\begin{definition}
For $k\geq 1$, we define $w_{k,2r}^{(g)}(x)$ to be the coefficient of order $\hbar^{g-2r}$ of $W_k^{(g)}(x,\dots,x)$. The index $r$ goes from $0$ to $\frac{g}{2}$ when $g$ is even and from $0$ to $\frac{g-1}{2}$ when $g$ is odd. In other words we have:
\begin{itemize}
\item When $g$ is even:
\bea W_1^{(g)}(x)&=&\hbar^g w_{1,0}^{(g)}(x)+ \hbar^{g-2}w_{1,2}^{(g)}(x) +\dots+ \hbar^{2}w_{1,g-2}^{(g)}(x)+ w_{1,g}^{(g)}(x)\cr
W_2^{(g)}(x,x)&=&\hbar^g w_{2,0}^{(g)}(x)+ \hbar^{g-2}w_{2,2}^{(g)}(x) +\dots+ \hbar^{2}w_{2,g-2}^{(g)}(x)+ w_{2,g}^{(g)}(x)\cr
&\dots&\cr
W_k^{(g)}(x,\dots,x)&=&\hbar^g w_{k,0}^{(g)}(x)+ \hbar^{g-2}w_{k,2}^{(g)}(x) +\dots+ \hbar^{2}w_{k,g-2}^{(g)}(x)+ w_{k,g}^{(g)}(x)\cr
\eea
\item When $g$ is odd:
\bea W_1^{(g)}(x)&=&\hbar^g w_{1,0}^{(g)}(x)+ \hbar^{g-2}w_{1,2}^{(g)}(x) +\dots+ \hbar w_{1,g-1}^{(g)}(x)\cr
W_2^{(g)}(x,x)&=&\hbar^g w_{2,0}^{(g)}(x)+ \hbar^{g-2}w_{2,2}^{(g)}(x) +\dots+ \hbar w_{2,g-1}^{(g)}(x)\cr
&\dots&\cr
W_k^{(g)}(x,\dots,x)&=&\hbar^g w_{k,0}^{(g)}(x)+ \hbar^{g-2}w_{k,2}^{(g)}(x) +\dots+ \hbar w_{k,g-1}^{(g)}(x)\cr
\eea
\end{itemize}
\end{definition} 

We can now state our generalized version of the conjecture:

\begin{conjecture}
For $n\geq 1$ and for $g\geq 0$ and $g$ \textbf{even}:
\bea \label{1} W_n^{(g)}(x,\dots,x)&=&\hbar^g\left(\frac{P_{n,1}^{(g)}(x)}{y^{5n+3g-7}}+\frac{P_{n,2}^{(g)}(x)}{y^{5n+3g-6}}\right)
+\hbar^{g-2}\left(\frac{P_{n,3}^{(g)}(x)}{y^{5n+3g-7}}+\frac{P_{n,4}^{(g)}(x)}{y^{5n+3g-6}}\right)\cr
&&+\dots+\hbar^2\left(\frac{P_{n,g-1}^{(g)}(x)}{y^{5n+3g-7}}+\frac{P_{n,g-2}^{(g)}(x)}{y^{5n+3g-6}}\right)+ \frac{P_{n,g+1}^{(g)}(x)}{y^{5n+3g-6}}\cr
\eea
where $\text{deg}(P_{g+1}^{(g)}(x))=2n+g-4$, for all $j$ \textbf{even}: $\text{deg}(P_{n,j}^{(g)})=2n+g-2$ and for all $j$ \textbf{odd}: $\text{deg}(P_{n,j}^{(g)})=2n+g-4$. Moreover, the polynomials are even or odd functions of $x$ according to their degree.\\
For $n\geq 1$ and for $g\geq 1$ and $g$ \textbf{odd}:
\bea \label{2} W_n^{(g)}(x,\dots,x)&=&\hbar^g\left(\frac{P_{n,1}^{(g)}(x)}{y^{5n+3g-7}}+\frac{P_{n,2}^{(g)}(x)}{y^{5n+3g-6}}\right)
+\hbar^{g-2}\left(\frac{P_{n,3}^{(g)}(x)}{y^{5n+3g-7}}+\frac{P_{n,4}^{(g)}(x)}{y^{5n+3g-6}}\right)+\dots\cr
&&+\hbar\left(\frac{P_{n,g}^{(g)}(x)}{y^{5n+3g-7}}+\frac{P_{n,g+1}^{(g)}(x)}{y^{5n+3g-6}}\right)\cr
\eea
where for all $j$ \textbf{even}: $\text{deg}(P_{n,j}^{(g)})=2n+g-2$ and for all $j$ \textbf{odd}: $\text{deg}(P_{n,j}^{(g)})=2n+g-3$. Moreover, the polynomials are even or odd functions of $x$ according to their degree. Moreover, the polynomials are even or odd functions of $x$ according to their degree.\\
Eventually we have at $x\to \infty$:
\beq \label{Asmp} W_n^{(g)}(x,\dots,x)=O\left(\frac{1}{x^{3n+2g-2}}\right)\eeq 
\end{conjecture}

Note that the last part of the conjecture is equivalent to prove that the leading coefficient of the polynomials $P_{n,j}^{(g)}$ and $P_{n,j+1}^{(g)}$ are opposite. Indeed, a trivial expansion at $x\to \infty$ of formulas \eqref{1} and \eqref{2} shows that we are two orders above the one claimed in \eqref{Asmp}. The first order is canceled by the condition on the leading coefficients of the polynomials and the second one vanishes trivially from the parity of the functions.

\medskip

\medskip

Unfortunately, we were not able to completely prove our conjecture with elementary means. Our best results are the following:

\begin{theorem} \label{Theo} The conjecture holds for $w_{1,0}^{(g)}(x)$, $w_{1,2}^{(g)}(x)$, $w_{2,0}^{(g)}(x)$ for any $g\geq 0$ as well as for any $w_{k,g}^{(g)}(x)$ with $k\geq 1$ and $g\geq 0$. In other words, we proved the conjecture of Witte and Forrester for the leading, subleading and last coefficients of the $\hbar$ expansion of $W_1^{(g)}(x)$.
\end{theorem}

Moreover we prove the asymptotic part of the conjecture at $x\to \infty$:

\begin{theorem}\label{Theo2} The asymptotic expansion at $x\to \infty$ of the functions $W_n^{(g)}(x,\dots,x)$ are of the form:
\beq W_n^{(g)}(x,\dots,x)=O\left(\frac{1}{x^{3n+2g-2}}\right)\eeq
\end{theorem}

The proofs of the two theorems are given in the next sections and in appendix \ref{AppendixA} we illustrate the validity of the conjecture with the computation of the first correlation functions.

\section{Proof of theorem \ref{Theo2}}

From the definition, it is clear that $\forall g\geq 1$, the functions $W_n^{(g)}(x,x_2,\dots,x_n)$ have a series expansion at $x\to \infty$ starting at least at $O\left(\frac{1}{x^2}\right)$. Let us start with the general loop equations \eqref{GeneralLoopExpanded} and let us take the asymptotic expansion at $x=\infty$. Since all correlation functions $W_n^{(g)}(x,x_1,\dots,x_{n-1})$ are of order at least $=O\left(\frac{1}{x^2}\right)$ at infinity and because $y(x)=x+O(1)$ we observe that the order $O(\left(\frac{1}{x}\right)$ only gets two contributions and we have:
\beq \label{BehaviorAtInfinity}\encadremath{\lim_{x\to \infty}x^2W_n^{(g)}(x,x_1,\dots,x_{n-1})=-\sum_{i=1}^{n-1}\partial_{x_{i}} W_{n-1}^{(g)}(x_1,\dots,x_{n-1}) }\eeq
Let us now focus on the expansion of $W_1^{(g)}(x)$ at $x=\infty$. The last equation for $n=2$ gives:
\beq \label{resint} \forall x_1\in \mathbb{C}\,:\,  W_2^{(g)}(x,x_1)\overset{x\to \infty}{=}\frac{\partial_{x_1}W_1^{(g)}(x_1)}{x^2}+ o\left(\frac{1}{x^2}\right)\eeq
The first loop equation can be rewritten as:
\beq \label{Need} y(x)W_1^{(g)}(x)=\sum_{p=1}^{g-1}W_1^{(p)}(x)W_1^{(g-p)}(x)+\hbar \partial_x W_1^{(g-1)}(x)+W_2^{(g-2)}(x,x)\eeq
We want to prove the expansion for $W_1^{(g)}(x)$ by induction on $g$. Let us assume that for $p\leq g-1$ we have $W_1^{(p)}(x)=O\left(\frac{1}{x^{2p+1}}\right)$ as wanted. From \eqref{resint} we get that $W_2^{(g-2)}(x,x)=O\left(\frac{\partial_{x}W_1^{(g-2)}(x)}{x^2}\right)=O\left(\frac{1}{x^{2g}}\right)$. All terms in the sum of \eqref{Need} are by induction of order $O\left(\frac{1}{x^{2g+2}}\right)$ while the derivative term is of order $O\left(\frac{1}{x^{2g}}\right)$. Dividing by $y(x)$gives that $W_1^{(g)}(x)$ is at least of order $O\left(\frac{1}{x^{2g+1}}\right)$. We also note that only terms in $\hbar \partial_x W_1^{(g-1)}(x)+W_2^{(g-2)}(x,x)$ contribute in the leading order of the expansion at $x\to \infty$ of $W_1^{(g)}(x)$. Since the induction initializes nicely for $W_1^{(1)}(x)$ and $W_1^{(2)}(x)$ (see explicit formulas) we get the following theorem:
For all $g\geq 1$ we have:
\bea W_1^{(g)}(x)&\overset{x\to \infty}{=}&O\left(\frac{1}{x^{2g+1}}\right)\cr
W_2^{(g)}(x,x_1)&\overset{x\to \infty}{=}&\frac{\partial_{x_1}W_1^{(g)}(x_1)}{x^2}+ o\left(\frac{1}{x^2}\right)\cr
\eea
Combining both results implies:
\beq W_2^{(g)}(x,x)\overset{x\to \infty}{=}O\left(\frac{1}{x^{2g+4}}\right)\eeq

From \eqref{BehaviorAtInfinity}, a trivial recursion on $n$ (with $g$ fixed) gives:
\beq \encadremath{ W_n^{(g)}(x,\dots,x)\overset{x\to \infty}{=}O\left(\frac{1}{x^{2g+3n-2}}\right) }\eeq
Indeed, at each step we gather a factor $\frac{1}{x^2}$ from the first variable and an additional derivative relatively to one variable that increases the degree by $1$. Therefore the exponent must be proportional to $3n$. The initialization of the induction is provided by $W_1^{(g)}$ and leads to the previous formula.

\section{Proof theorem \ref{Theo}}

\subsection{The case of $w_{k,g}^{(g)}(x)$}

The terms $w_{k,g}^{(g)}(x)$ correspond to the limit when $\hbar\to 0$ of the correlation functions:
\beq w_{k,g}^{(g)}(x)=\lim_{\hbar\to 0} W_k^{(g)}(x,\dots,x)\eeq
These terms correspond to the hermitian case for which many results are known. In particular, the structure proposed in the conjecture has been known for a long time and can be easily recovered by standard results regarding the topological recursion described in \cite{EO} for the spectral curve $y^2=x^2-4T$. The main challenge is thus to prove that the structure remains valid for higher order in $\hbar$.

\subsection{The $\hbar^g$ order in $W_1^{(g)}(x)$}

First we prove the properties for the $\hbar^g$ order of $W_1^{(g)}(x)$, that is to say with our notation for $w_{1,0}^{(g)}(x)$. The strategy is the following: we observe that the loop equation \eqref{ExpandedFirstLoop} projects into the highest order in $\hbar$ like:
\beq \label{w1gx}w_{1,0}^{(g)}(x)=\frac{1}{y(x)}\left[\partial_x w_{1,0}^{(g-1)}(x)+\sum_{p=1}^{g-1}w_{1,0}^{(p)}(x)w_{1,0}^{(g-p)}(x)\right]\eeq
In particular, note that the term $W_2^{(g-2)}(x,x)$ cannot provide any contribution here because its degree in $\hbar$ is at most $\hbar^{g-2}$. Now the previous equation gives a recursive way to compute $w_{1,0}^{(g)}(x)$ from the knowledge of $w_{1,0}^{(1)}(x)=\frac{1}{2}\left(\frac{1}{y(x)}-\frac{x}{y(x)^2}\right)$. Let us prove by induction on $g$ that the properties presented in the theorem hold for $w_{1,0}^{(g)}(x)$. First we observe that it holds for $w_{1,0}^{(1)}(x)$ thus initializing the induction (note that $W_1^{(0)}(x)$ never appears in the loop equations in the double sum so we do not need it to perform the recursion). Then inserting the desired form of $w_{1,0}^{(k)}(x)$ with $k\leq g-1$ into \eqref{w1gx} leads to:

\bea \label{Po} P_{1,1}^{(g)}(x)&=&-x(3g-5)P_{1,1}^{(g-1)}(x)+(x^2-4T)P_{1,1}^{(g-1)'}(x)+\sum_{p=1}^{g-1}P_{1,1}^{(p)}(x)P_{1,2}^{(g-p)}(x)+P_{1,2}^{(p)}(x)P_{1,1}^{(g-p)}(x)\cr
P_{1,2}^{(g)}(x)&=&-x(3g-4)P_{1,2}^{(g-1)}(x)+(x^2-4T)P_{1,2}^{(g-1)'}(x)+\sum_{p=1}^{g-1}P_{1,2}^{(p)}(x)P_{1,2}^{(g-p)}(x)+P_{1,1}^{(p)}(x)P_{1,1}^{(g-p)}(x)\cr
\eea

It is then straightforward to see that the claimed degrees for $P_{1,1}^{(g)}(x)$ and $P_{1,2}^{(g)}(x)$ match properly with the r.h.s. to give respectively $g-1$ and $g$. The parity of the polynomials also follows directly from the one arising in the r.h.s. (one needs to split into cases $g$ even or $g$ odd to write it properly but no problem arises here). The claim regarding the leading coefficients of the polynomials is a little more subtle. We denote by $p_{1,1}^{(g)}$ and $p_{1,2}^{(g)}$ the leading coefficients of respectively  $P_{1,1}^{(g)}$ and $P_{1,2}^{(g)}$. Looking at the leading coefficients of \eqref{Po} leads to:
\bea p_{1,1}^{(g)}&=&-(3g-5)p_{1,1}^{(g-1)}+(g-1)p_{1,1}^{(g-1)}+\sum_{p=1}^{g-1}p_{1,1}^{(p)}p_{1,2}^{(g-p)}+p_{1,2}^{(p)}p_{1,1}^{(g-p)}\cr
p_{1,2}^{(g)}&=&-(3g-4)p_{1,2}^{(g-1)}+g p_{1,2}^{(g-1)}+\sum_{p=1}^{g-1}p_{1,2}^{(p)}p_{1,2}^{(g-p)}+p_{1,1}^{(p)}p_{1,1}^{(g-p)}\cr 
\eea
If we assume by induction that $p_{1,1}^{(k)}+p_{1,2}^{(k)}=0$ for all $k\leq g-1$ then summing the last two equations gives $p_{1,1}^{(g)}+p_{1,2}^{(g)}=0$. Hence we have proved here that the main theorem holds for $w_{1,0}^{(g)}(x)$.
 
\subsection{The $\hbar^g$ order in $W_2^{(g)}(x,x)$}

In order to prove the results for $w_{2,0}^{(g)}(x)$ we first need to take the general loop equation \eqref{GeneralLoopExpanded} at coinciding points $x_1=x_2=\dots=x_n$. We know from their definition that the correlations functions $W_n^{(g)}(x_1,\dots,x_n)$ are symmetric functions in $x_1,\dots,x_n$ and that they are regular at $x_1=\dots=x_n$. We get:
 
\bea \label{GeneralLoopExpandedxx} &&y(x)W_n^{(g)}(x,\dots,x)=\frac{\hbar}{n} \partial_x\left(W_n^{(g-1)}(x,\dots,x)\right)+W_{n+1}^{(g-2)}(x,x,\dots,x)\cr
&&+\sum_{J\subseteq I}'\sum_{p=0}^gW_{|J|+1}^{(p)}(x,\dots,x)W_{n-|J|}^{(g-p)}(x,\dots,x)+\frac{(n-1)}{2}\left(\partial_1^2W_{n-1}^{(g)}\right)(x,\dots,x)\cr
\eea
where the notation $\left(\partial_1^2W_{n-1}^{(g)}\right)(x,\dots,x)$ means that we must take twice the derivatives of $(x_1,\dots,x_n)\mapsto W_{n-1}^{(g)}(x_1,\dots,x_n)$ in its first variable $x_1$ and then evaluate it at coinciding points $x_1=\dots=x_n=x$. In particular for $n=2$, \eqref{GeneralLoopExpandedxx} leads to:

\beq \label{Loopw2gx}  y(x)W_2^{(g)}(x,x)=\frac{\hbar}{2}\partial_x\left(W_2^{(g-1)}(x,x)\right)+W_3^{(g-2)}(x,x,x)+2\sum_{p=0}^{g-1}W_2^{(p)}(x,x)W_1^{(g-p)}(x) + \frac{1}{2}W_1^{(g)''}(x) \eeq

Taking the coefficient in $\hbar^g$ gives:

\beq \label{Loopw2gxx}  y(x)w_{2,0}^{(g)}(x)=\frac{1}{2}\partial_x\left(w_{2,0}^{(g-1)}(x)\right)+2\sum_{p=0}^{g-1}w_{2,0}^{(p)}(x)w_{1,0}^{(g-p)}(x) + \frac{1}{2}w_{1,0}^{(g)''}(x) \eeq

Since we know the properties for $w_{1,0}^{(g)}(x)$ from the last subsection, we get a recursive relation that determines $w_{2,0}^{(g)}(x)$ from the lower cases $w_{2,0}^{(k)}(x)$ with $k\leq g-1$. We note also that the properties stated in the theorem hold for $W_2^{(0)}(x,x)=\frac{T}{y(x)^4}$ hence initializing our induction. Inserting the desired form and the knowledge about $w_{1,0}^{(g)}(x)$ into \eqref{Loopw2gxx} gives: (we mention here that according to \ref{Derivy} we have the following observations: $y(x)^2=x^2-4T$, $y'(x)=\frac{x}{y(x)}$ and $y''(x)=-\frac{4T}{y(x)^3}$)

\bea \label{Poo} P_{2,2}^{(g)}(x)&=&-\frac{(3g+1)x}{2}P_{2,2}^{(g-1)}(x)+\frac{x^2-4T}{2}P_{2,2}^{(g-1)'}(x)\cr
&&+2\sum_{p=0}^{g-1}\left(P_{2,2}^{(p)}(x)P_{1,2}^{(g-p)}(x)+(x^2-4T)P_{2,1}^{(p)}(x)P_{1,1}^{(g-p)}(x)\right) \cr
&&+\frac{1}{2}g(3g-1)(4+3x^2)P_{1,2}^{(g)}(x)-(3g-1)x(x^2-4T)P_{1,2}^{(g)'}(x)+\frac{1}{2}(x^2-4T)P_{1,2}^{(g)''}(x)\cr
P_{2,1}^{(g)}(x)&=&-\frac{3gx}{2}P_{2,1}^{(g-1)}(x)+\frac{x^2-4T}{2}P_{2,1}^{(g-1)'}(x)+2\sum_{p=0}^{g-1}\left(P_{2,1}^{(p)}(x)P_{1,2}^{(g-p)}(x)+P_{2,2}^{(p)}(x)P_{1,1}^{(g-p)}(x)\right) \cr
&&+\frac{1}{2}(3g-2)(4g+(3g-1)x^2)P_{1,1}^{(g)}(x)-(3g-2)x(x^2-4T)P_{1,1}^{(g)'}(x)\cr
&&+\frac{1}{2}(x^2-4T)P_{1,1}^{(g)''}(x)
\eea 

We remind here that a prime means a derivative relatively to $x$. Similarly to the previous case, it is straightforward to observe that the degree and parity properties extend from the cases $k\leq g-1$ to $g$ by using the results for $P_{1,1}^{(p)}(x)$ and $P_{1,2}^{(p)}(x)$. The leading coefficients are again a little more subtle. Indeed the leading coefficients in $x$ of \eqref{Poo} gives: 
\bea p_{2,2}^{(g)}&=& -\frac{(3g+1)}{2}p_{2,2}^{(g-1)}+\frac{g+1}{2}p_{2,2}^{(g-1)}+2\sum_{p=0}^{g-1}\left(p_{2,2}^{(p)}p_{1,2}^{(g-p)}+p_{2,1}^{(p)}p_{1,1}^{(g-p)}\right) \cr
&&+\frac{3}{2}g(3g-1)p_{1,2}^{(g)}-(3g-1)g p_{1,2}^{(g)}+\frac{g(g-1)}{2}p_{1,2}^{(g)}\cr
p_{2,1}^{(g)}&=&-\frac{3g}{2}p_{2,1}^{(g-1)}+\frac{g}{2}p_{2,1}^{(g-1)}+2\sum_{p=0}^{g-1}\left(p_{2,1}^{(p)}p_{1,2}^{(g-p)}+p_{2,2}^{(p)}p_{1,1}^{(g-p)}\right)\cr
&&+\frac{(3g-2)(3g-1)}{2}p_{1,1}^{(g)}-(3g-2)(g-1)p_{1,1}^{(g)}+\frac{(g-1)(g-2)}{2}p_{1,1}^{(g)}\cr
\eea
Then one observes the identity :
\beqq \frac{3}{2}g(g-1)-(3g-1)g+\frac{1}{2}g(g-1)=\frac{1}{2}(3g-2)(3g-1)-(3g-2)(g-1)+\frac{1}{2}(g-1)(g-2)\eeqq
so that the last lines of each quantity are opposite. Hence the induction gives $p_{2,1}^{(g)}+p_{2,2}^{(g)}=0$ concluding the proof for the $w_{2,0}^{(g)}(x)$.

\subsection{The $\hbar^{g-2}$ order in $W_1^{(g)}(x)$}

The loop equation for $n=1$ is given by:
\beq y(x)W_1^{(g)}(x)=\sum_{p=1}^{g-1}W_1^{(p)}(x)W_1^{(g-p)}(x)+\hbar W_1^{(g-1)'}(x)+W_2^{(g-2)}(x,x)\eeq
Taking order $\hbar^{g-2}$ in the previous equation gives:
\beq \label{w12x} w_{1,2}^{(g)}(x)=\frac{1}{y(x)}\left[2\sum_{p=2}^{g-1}w_{1,2}^{(p)}w_{1,0}^{(g-p)}(x)+w_{1,2}^{(g-1)'}(x)+w_{2,0}^{(g-2)}(x)\right]\eeq
From the last section, we know that the desired properties hold for $w_{1,0}^{(k)}(x)$ and $w_{2,0}^{(k)}(x)$. Therefore the previous equation gives us a recursive way to get $w_{1,2}^{(g)}(x)$ from the lower cases $w_{1,2}^{(k)}(x)$ with $k\leq g-1$. The recursion holds for $w_{1,2}^{(1)}(x)=0$ as well as $w_{1,2}^{(2)}(x)=w_{2,0}^{(0)}(x)$ that are known cases. A straightforward computation gives:
\bea P_{1,3}^{(g)}(x)&=&2\sum_{p=2}^{g-1}\left( P_{1,3}^{(p)}(x)P_{1,2}^{(g-p)}(x)+ P_{1,4}^{(p)}(x)P_{1,1}^{(g-p)}(x)\right) +(x^2-4T)P_{1,3}^{(g-1)'}(x)\cr
&&+(3g-5)xP_{1,3}^{(g-1)}(x)+P_{2,1}^{(g-2)}(x)\cr
P_{1,4}^{(g)}(x)&=& 2\sum_{p=2}^{g-1}\left( P_{1,4}^{(p)}(x)P_{1,2}^{(g-p)}(x)+ (x^2-4T)P_{1,2}^{(p)}(x)P_{1,1}^{(g-p)}(x)\right) +(x^2-4T)P_{1,4}^{(g-1)'}(x)\cr
&&+(3g-4)xP_{1,4}^{(g-1)}(x)+P_{2,2}^{(g-2)}(x)\cr
\eea
From the last set of equations, it is then easy to observe from the knowledge of $P_{1,1}^{(k)}(x)$, $P_{1,2}^{(k)}(x)$, $P_{2,1}^{(k)}(x)$ and $P_{2,2}^{(k)}(x)$ that an easy recursion on $g$ will lead to the fact that $P_{1,3}^{(g)}(x)$ and $P_{1,4}^{(g)}(x)$ satisfy the expected conditions on their degree, parity and leading coefficients as stated in the conjecture. The induction goes in the same spirit as before and presents no difficulty. Therefore we conclude that the desired properties hold for $w_{1,2}^{(g)}(x)$.

\section{Limitation of our strategy}

The strategy involved in our previous proofs is to use the loop equations only at coinciding points $x_1=\dots=x_n=x$ and to use a recursive way to prove our conjecture. In general the loop equations \eqref{GeneralLoopExpanded} at coinciding points are:

\bea \label{LoopCoinciding} &&y(x)W_n^{(g)}(x,\dots,x)=\frac{\hbar}{n}\partial_x \left(W_n^{(g-1)}(x,\dots,x)\right)+W_{n+1}^{(g-2)}(x,x,\dots,x)\cr
&&+\sum_{j=0}^{n-1'}\sum_{p=0}^g \binom{n-1}{j} W_{|J|+1}^{(p)}(x,\dots,x)W_{n-|J|}^{(g-p)}(x,\dots,x)+\frac{n-1}{2}\left(\partial_1^2 W_{n-1}^{(g)}\right)_{|(x,\dots,x)}
\eea 
In the double sum, we exclude as usual the terms $(j,p)=(0,0)$ and $(n-1,g)$. Note here that we have used the symmetry of the functions $W_{n}^{(g-1)}(x_1,\dots,x_{n-1})$ in order to rewrite the term involving $\hbar$. At this point it is tempting to define $f_n^{(g)}(x)=W_n^{(g)}(x,\dots,x)$ and hope that previous equation will lead to a nice induction. Unfortunately this is not true when $n\geq 3$ because the term $\left(\partial_1^2 W_{n-1}^{(g)}\right)_{|(x,\dots,x)}$ cannot be rewritten easily with $f_n^{(g)}(x)$. Indeed, the second derivative of $f_{n-1}^{(g)}(x)$ would imply terms like $\left(\partial_i\partial_j W_{n-1}^{(g)}\right)_{|(x,\dots,x)}$ that become problematic when $i\neq j$. For $n=1$ and $n=2$, these terms are no longer problematic and we could hope to derive results for these cases (that in particular recover the conjecture of Witte and Forrester). Unfortunately, equation \eqref{LoopCoinciding} contains the term $W_{n+1}^{(g-2)}(x,x,\dots,x)$ that increases the value of $n$ thus prohibiting a simple recursive approach for every order in $\hbar$. However as we proved in this article the conjecture holds for highest and lowest orders in $\hbar$ and there is little doubt that it should hold for the middle ones. A possible approach could be to consider the full correlation functions $W_n^{(g)}(x_1,\dots,x_n)$ and propose a general form that we could insert into the loop equations and that we could prove by induction. This strategy looks tedious because the proposed form should be precise enough to contain our conjecture but general enough to be proved by induction. Moreover the computations proposed in appendix \ref{AppendixA} show that the coinciding point limit is very singular and prevented us to guess a general formula for $W_n^{(g)}(x_1,\dots,x_n)$. Another possible approach could be to use the expansion at $x\to \infty$ and insert it in a clever way into the loop equations to get information about the problematic term $\left(\partial_1^2 W_{n-1}^{(g)}\right)_{|(x,\dots,x)}$. Eventually a last possible way to prove the conjecture could be to propose a formula for every derivative $\left(\partial_1^{i_1}\dots\partial_n^{i_n} W_{n}^{(g)} \right)_{|(x,\dots,x)}$ and prove them by induction. In this approach, the difficult step is no longer becomes to go from $(k,p)$ with $k<n$ and $p<g$ to $(n,g)$ for which \eqref{LoopCoinciding} applies nicely but to find a way to get the derivatives of $W_n^{(g)}(x,\dots,x)$ from the loop equations. The author would be very happy to work with anyone interested with this problem.

\begin{appendices}

\section{\label{AppendixA}Illustration of the conjecture: computation of the first correlation functions}

In this section we illustrate our conjecture with the computation of the first correlation functions $W_n^{(g)}(x_1,\dots,x_n)$ as well as their limit at coinciding points $x_1=\dots=x_n=x$. We find:

\medskip
\underline{$W_2^{(1)}(x_1,x_2)$}:
\medskip

\bea \label{W21}W_2^{(1)}(x_1,x_2)&=&\frac{\hbar}{2}\Big[\frac{x_1x_2+4T}{y_1^3y_2^3}-\frac{x_1^2x_2^2+4Tx_1^2-4Tx_1x_2-3x_1x_2^3-32T^2+16Tx_2^2}{(x_1-x_2)^3y_1y_2^4}\cr
&&+\frac{x_1^2x_2^2+4Tx_2^2-4Tx_1x_2-3x_1^3x_2-32T^2+16Tx_1^2}{(x_1-x_2)^3y_1^4y_2}\Big]\cr
\eea
which non trivially gives:
\beq W_2^{(1)}(x,x)=\hbar\left(-\frac{x(x^2+18T)}{2y^7}+\frac{x^2+4T}{2y^6}\right)\eeq
The limit at coinciding points is non trivial and comes from the fact that:
\beq P(x_1,x_2)=x_1^2x_2^2+4Tx_2^2-4Tx_1x_2-3x_1^3x_2-32T^2+16Tx_1^2 \Rightarrow P(x,x)=-2y(x)^4\eeq

\medskip
\underline{$W_2^{(2)}(x_1,x_2)$}:
\medskip

We have also:
\bea W_2^{(2)}(x_1,x_2)&=&\frac{A(x_1,x_2)}{y_1^7y_2^7}+ \hbar^2\Big[ \frac{P(x_1,x_2)}{y_1^4y_2^4(x_1-x_2)^4}+ \frac{Q(x_1,x_2)}{y_1^3y_2^6(x_1-x_2)^3}\cr
&&- \frac{Q(x_2,x_1)}{y_1^6y_2^3(x_1-x_2)^3}+\frac{S(x_1,x_2)}{y_1^7y_2^7(x_1-x_2)^4}\Big]\cr
\eea
with:
\bea A(x_1,x_2)&=&T\Big(5\,{x_{{1}}}^{5}x_{{2}}+5\,x_{{1}}{x_{{2}}}^{5}+4\,{x_{{1}}}^{4}{x_{{2}}}^{2}+4\,{x_{{1}}}^{2}{x_{{2}}}^{4}+3\,{x_{{1}}}^{3}{x_{{2}}}^{3} \Big)\cr
&&+T^2\Big(-52\,{x_{{1}}}^{3}x_{{2}}-52\,x_{{1}}{x_{{2}}}^{3}-52\,{x_{{1}}}^{2}{x_{{2}}}^{2}+4\,{x_{{1}}}^{4}+4\,{x_{{2}}}^{4}\Big)\cr
&&+T^3\Big(-16\,{x_{{1}}}^{2}-16\,{x_{{2}}}^{2}+208\,x_{{1}}x_{{2}} \Big)+320\,T^4\cr
P(x_1,x_2)&=&\frac{3}{2}\,{x_{{1}}}^{4}{x_{{2}}}^{2}+\frac{3}{2}\,{x_{{1}}}^{2}{x_{{2}}}^{4}-6\,{x_{{1}}}^{3}{x_{{2}}}^{3}\cr
&&+ T\Big( 6\,{x_{{1}}}^{4}+6\,{x_{{2}}}^{4}-8\,x_{{1}}{x_{{2}}}^{3}-8\,{x_{{1}}}^{3}x_{{2}} +40\,{x_{{1}}}^{2}{x_{{2}}}^{2}\Big)\cr
&&+ T^2\Big(-88\,{x_{{1}}}^{2}-88\,{x_{{2}}}^{2}+32\,x_{{1}}x_{{2}} \Big) +192\,T^3\cr
Q(x_1,x_2)&=&-3\,{x_{{1}}}^{4}{x_{{2}}}^{2}+7\,{x_{{1}}}^{3}{x_{{2}}}^{3}-\frac{9}{2}\,{x_{{1}}}^{2}{x_{{2}}}^{4}+\frac{3}{2}\,x_{{1}}{x_{{2}}}^{5}\cr
&&+T\Big(-4\,{x_{{1}}}^{4}+4\,{x_{{1}}}^{3}x_{{2}}+12\,{x_{{1}}}^{2}{x_{{2}}}^{2}+8\,{x_{{2}}}^{4}-32\,x_{{1}}{x_{{2}}}^{3} \Big)\cr
&&+T^2\Big(-24\,x_{{1}}x_{{2}}+24\,{x_{{1}}}^{2}+48\,{x_{{2}}}^{2} \Big) -64\,T^3\cr
S(x_1,x_2)&=&\frac{23}{2}\,{x_{{1}}}^{7}{x_{{2}}}^{5}+\frac{23}{2}\,{x_{{1}}}^{5}{x_{{2}}}^{7}-10\,{x_{{1}}}^{8}{x_{{2}}}^{4}-10\,{x_{{1}}}^{4}{x_{{2}}}^{8}\cr
&&+3\,{x_{{1}}}^{9}{x_{{2}}}^{3}+3\,{x_{{1}}}^{3}{x_{{2}}}^{9}-6\,{x_{{1}}}^{6}{x_{{2}}}^{6}\cr
&&+ T\Big( -36\,{x_{{1}}}^{2}{x_{{2}}}^{8}-36\,{x_{{1}}}^{8}{x_{{2}}}^{2}+128\,{x_{{1}}}^{4}{x_{{2}}}^{6}+128\,{x_{{1}}}^{6}{x_{{2}}}^{4}-7\,{x_{{1}}}^{7}{x_{{2}}}^{3}-7\,{x_{{1}}}^{3}{x_{{2}}}^{7}\cr
&&+13\,{x_{{1}}}^{9}x_{{2}}+13\,x_{{1}}{x_{{2}}}^{9}-268\,{x_{{1}}}^{5}{x_{{2}}}^{5} \Big) \cr
&&+T^2 \Big( -156\,{x_{{1}}}^{7}x_{{2}}-156\,x_{{1}}{x_{{2}}}^{7}+388\,{x_{{1}}}^{6}{x_{{2}}}^{2}+388\,{x_{{1}}}^{2}{x_{{2}}}^{6}+4\,{x_{{1}}}^{8}+4\,{x_{{2}}}^{8}\cr
&&+284\,{x_{{1}}}^{3}{x_{{2}}}^{5}+284\,{x_{{1}}}^{5}{x_{{2}}}^{3}-320\,{x_{{1}}}^{4}{x_{{2}}}^{4} \Big)\cr
&&+ T^3\Big( -3376\,{x_{{1}}}^{4}{x_{{2}}}^{2}-3376\,{x_{{1}}}^{2}{x_{{2}}}^{4}+16\,x_{{1}}{x_{{2}}}^{5}-16\,{x_{{2}}}^{6}-16\,{x_{{1}}}^{6}+16\,{x_{{1}}}^{5}x_{{2}} +2912\,{x_{{1}}}^{3}{x_{{2}}}^{3}\Big)\cr
&&+ T^4\Big( 12160\,{x_{{1}}}^{2}{x_{{2}}}^{2}+3392\,{x_{{1}}}^{4}+3392\,{x_{{2}}}^{4}-3712\,x_{{1}}{x_{{2}}}^{3}-3712\,{x_{{1}}}^{3}x_{{2}} \Big)\cr
&& + T^5\Big( -10240\,{x_{{1}}}^{2}-10240\,{x_{{2}}}^{2}+2048\,x_{{1}}x_{{2}} \Big) +12288\,T^6\cr
\eea
Note that $A(x,x)\propto y(x)^4$, $P(x,x)\propto y(x)^6$, $Q(x,x)\propto y(x)^6$ and $S(x,x)\propto y(x)^{12}$ 

\medskip
\underline{$W_3^{(1)}(x_1,x_2,x_3)$}:
\medskip

\bea &&W_3^{(1)}(x_1,x_2,x_3)=\hbar\Big[ \frac{Q_{111}(x_1,x_2,x_3)}{y_1^5y_2^5y_3^5}+\frac{Q_{110}(x_1,x_2,x_3)}{y_1^3y_2^3y_3^6(x_1-x_3)^3(x_2-x_3)^3}\cr
&&+\frac{Q_{101}(x_1,x_2,x_3)}{y_1^3y_2^6y_3^3(x_1-x_2)^3(x_2-x_3)^3}+\frac{Q_{011}(x_1,x_2,x_3)}{y_1^6y_2^3y_3^3(x_1-x_2)^3(x_1-x_3)^3}\cr
\eea
with polynomials $Q$ given by:
\bea Q_{111}&=&{x_{{1}}}^{2}{x_{{2}}}^{3}{x_{{3}}}^{3}+{x_{{1}}}^{3}{x_{{2}}}^{3}{x_{{3}}}^{2}+{x_{{1}}}^{3}{x_{{2}}}^{2}{x_{{3}}}^{3}\cr
&&+T\Big(8\,{x_{{1}}}^{2}{x_{{2}}}^{2}{x_{{3}}}^{2}+2\,{x_{{1}}}^{2}{x_{{2}}}^{3}x_{{3}}+2\,x_{{1}}{x_{{2}}}^{3}{x_{{3}}}^{2}+2\,{x_{{1}}}^{3}x_{{2}}{x_{{3}}}^{2}+2\,{x_{{1}}}^{2}x_{{2}}{x_{{3}}}^{3}\cr
&&+2\,x_{{1}}{x_{{2}}}^{2}{x_{{3}}}^{3}+2\,{x_{{1}}}^{3}{x_{{2}}}^{2}x_{{3}}+2\,{x_{{1}}}^{3}{x_{{3}}}^{3}+2\,{x_{{1}}}^{3}{x_{{2}}}^{3}+2\,{x_{{2}}}^{3}{x_{{3}}}^{3}\Big)\cr
&&+T^2\Big(-8\,{x_{{2}}}^{2}{x_{{3}}}^{2}-8\,{x_{{1}}}^{2}{x_{{2}}}^{2}-8\,{x_{{1}}}^{2}{x_{{3}}}^{2}-32\,x_{{1}}{x_{{2}}}^{2}x_{{3}}-32\,x_{{1}}x_{{2}}{x_{{3}}}^{2}-32\,{x_{{1}}}^{2}x_{{2}}x_{{3}}\cr
&&-32\,x_{{1}}{x_{{3}}}^{3}-32\,x_{{1}}{x_{{2}}}^{3}-32\,{x_{{1}}}^{3}x_{{3}}-32\,x_{{2}}{x_{{3}}}^{3}-32\,{x_{{2}}}^{3}x_{{3}}-32\,{x_{{1}}}^{3}x_{{2}}\Big)\cr
&&+T^3\Big(224\,x_{{1}}x_{{3}}+224\,x_{{2}}x_{{3}}+224\,x_{{1}}x_{{2}}-64\,{x_{{1}}}^{2}-64\,{x_{{2}}}^{2}-64\,{x_{{3}}}^{2}\Big)+640\,{T}^{4}\cr
\eea

\bea Q_{110}&=&+2\,{x_{{1}}}^{3}{x_{{2}}}^{4}{x_{{3}}}^{4}+2\,{x_{{1}}}^{4}{x_{{2}}}^{3}{x_{{3}}}^{4} -3\,{x_{{1}}}^{3}{x_{{2}}}^{3}{x_{{3}}}^{5}-{x_{{1}}}^{4}{x_{{2}}}^{4}{x_{{3}}}^{3}\cr
&&+T\Big(-18\,{x_{{2}}}^{3}{x_{{3}}}^{6}-18\,{x_{{1}}}^{3}{x_{{3}}}^{6}-6\,x_{{1}}{x_{{3}}}^{8}+48\,x_{{1}}{x_{{2}}}^{3}{x_{{3}}}^{5}+24\,{x_{{1}}}^{4}{x_{{2}}}^{3}{x_{{3}}}^{2}-18\,x_{{1}}{x_{{2}}}^{4}{x_{{3}}}^{4}\cr
&&+6\,{x_{{1}}}^{2}{x_{{2}}}^{3}{x_{{3}}}^{4}-52\,{x_{{1}}}^{3}{x_{{2}}}^{3}{x_{{3}}}^{3}-6\,x_{{2}}{x_{{3}}}^{8}-12\,{x_{{1}}}^{4}{x_{{2}}}^{4}x_{{3}}+36\,{x_{{1}}}^{2}{x_{{2}}}^{2}{x_{{3}}}^{5}-18\,{x_{{1}}}^{4}x_{{2}}{x_{{3}}}^{4}\cr
&&+48\,{x_{{1}}}^{3}x_{{2}}{x_{{3}}}^{5}+6\,{x_{{1}}}^{3}{x_{{2}}}^{2}{x_{{3}}}^{4}-48\,{x_{{1}}}^{2}x_{{2}}{x_{{3}}}^{6}-48\,x_{{1}}{x_{{2}}}^{2}{x_{{3}}}^{6}+6\,{x_{{2}}}^{4}{x_{{3}}}^{5}-12\,{x_{{1}}}^{2}{x_{{2}}}^{4}{x_{{3}}}^{3}\cr
&&+24\,{x_{{1}}}^{3}{x_{{2}}}^{4}{x_{{3}}}^{2}+28\,x_{{1}}x_{{2}}{x_{{3}}}^{7}-12\,{x_{{1}}}^{4}{x_{{2}}}^{2}{x_{{3}}}^{3}+18\,{x_{{1}}}^{2}{x_{{3}}}^{7}+18\,{x_{{2}}}^{2}{x_{{3}}}^{7}+6\,{x_{{1}}}^{4}{x_{{3}}}^{5}\Big)\cr
&&+T^2\Big(88\,x_{{2}}{x_{{3}}}^{6}+48\,{x_{{2}}}^{4}{x_{{3}}}^{3}-32\,{x_{{3}}}^{7}+48\,{x_{{1}}}^{4}{x_{{3}}}^{3}-72\,{x_{{2}}}^{2}{x_{{3}}}^{5}-48\,{x_{{2}}}^{3}{x_{{3}}}^{4}+48\,{x_{{1}}}^{2}{x_{{2}}}^{4}x_{{3}} \cr
&&+48\,{x_{{1}}}^{4}{x_{{2}}}^{2}x_{{3}}-48\,x_{{1}}{x_{{2}}}^{4}{x_{{3}}}^{2}-48\,{x_{{1}}}^{4}x_{{2}}{x_{{3}}}^{2}+48\,x_{{1}}{x_{{2}}}^{3}{x_{{3}}}^{3}-144\,{x_{{1}}}^{2}{x_{{2}}}^{2}{x_{{3}}}^{3}\cr
&&+168\,x_{{1}}{x_{{2}}}^{2}{x_{{3}}}^{4}+168\,{x_{{1}}}^{2}x_{{2}}{x_{{3}}}^{4}+48\,{x_{{1}}}^{3}x_{{2}}{x_{{3}}}^{3}-288\,x_{{1}}x_{{2}}{x_{{3}}}^{5}\cr
&&-72\,{x_{{1}}}^{2}{x_{{3}}}^{5}+88\,x_{{1}}{x_{{3}}}^{6}-48\,{x_{{1}}}^{3}{x_{{3}}}^{4}\Big)  \cr
&&+T^3\Big(96\,{x_{{1}}}^{2}{x_{{3}}}^{3}-96\,{x_{{1}}}^{3}{x_{{3}}}^{2}-96\,{x_{{1}}}^{3}{x_{{2}}}^{2}+96\,{x_{{2}}}^{2}{x_{{3}}}^{3}-96\,{x_{{1}}}^{2}{x_{{2}}}^{3}-96\,{x_{{2}}}^{3}{x_{{3}}}^{2}\cr
&&-32\,{x_{{1}}}^{4}x_{{3}}-32\,{x_{{2}}}^{4}x_{{3}}-32\,{x_{{1}}}^{4}x_{{2}}-32\,x_{{1}}{x_{{2}}}^{4}+64\,{x_{{1}}}^{3}x_{{2}}x_{{3}}+64\,x_{{1}}{x_{{2}}}^{3}x_{{3}}\cr
&&-256\,x_{{1}}x_{{2}}{x_{{3}}}^{3}+224\,x_{{2}}{x_{{3}}}^{4}+224\,x_{{1}}{x_{{3}}}^{4}\Big) \cr
&&+T^4\Big( 384\,x_{{1}}{x_{{2}}}^{2}-384\,x_{{2}}^{2}x_{{3}}+384\,x_{{2}}{x_{{3}}}^{2} +384\,x_{{1}}{x_{{3}}}^{2}+384\,{x_{{1}}}^{2}x_{{2}}-384\,{x_{{1}}}^{2}x_{{3}}\cr
&&-1024\,{x_{{3}}}^{3}+256\,{x_{{1}}}^{3}+256\,{x_{{2}}}^{3}-256\,x_{{1}}x_{{2}}x_{{3}}\Big) \cr
&&+T^5\Big(2048\,x_{{3}}-1024\,x_{{1}}-1024\,x_{{2}}\Big)  \cr
\eea
and
\beq Q_{101}(x_1,x_2,x_3)=-Q_{110}(x_1,x_3,x_2) \,\,\text{ and } Q_{011}(x_1,x_2,x_3)=Q_{110}(x_3,x_2,x_1)\eeq

$Q_{111}$ is a symmetric polynomial and we have:
\bea Q_{111}(x,x,x)&=&y(x)^4(3x^4+50Tx^2+40T^2)\cr
 \frac{Q_{110}(x,x_2,x_3)}{(x-x_3)^3(x_2-x_3)^3}&\underset{x_2,x_3\to x}{\to}& 2x(216T^2-122x^2+21x^4)\cr
\eea
We can obtain similar limits for $Q_{101}$ and $Q_{011}$ so we find:
\beq W_3^{(1)}(x,x,x)=\hbar\left[\frac{3x^4+50Tx^2+40T^2}{y^{11}}-\frac{x(3x^4+160Tx^2+354T^2)}{y^{12}}\right]\eeq

We checked our conjecture for all correlation functions required to obtain $W_1^{(5)}(x)$. As one can see from the previous example, taking coinciding points $x_1=\dots=x_n=x$ is highly non trivial and increases the power of $y(x)$ at the denominator quite substantially.

\end{appendices}

\end{document}